\documentclass[conference]{IEEEtran}

\usepackage{amsmath}
\usepackage{color}
\usepackage{graphicx}
\usepackage{hyperref}
\usepackage{url}

\begin{document}

\title{Temperature Dependence of Magnetic Properties of an 18-nm-thick
  YIG Film Grown by Liquid Phase Epitaxy: Effect of a Pt Overlayer}

\author{\IEEEauthorblockN{Nathan Beaulieu\IEEEauthorrefmark{1}$^,$\IEEEauthorrefmark{2},
Nelly Kervarec\IEEEauthorrefmark{3},
Nicolas Thiery\IEEEauthorrefmark{4},
Olivier Klein\IEEEauthorrefmark{4}, 
Vladimir Naletov\IEEEauthorrefmark{2}$^,$\IEEEauthorrefmark{4}$^,$\IEEEauthorrefmark{5},\\
Herv\'e Hurdequint\IEEEauthorrefmark{2},
Gr\'egoire de Loubens\IEEEauthorrefmark{2},
Jamal Ben Youssef\IEEEauthorrefmark{1}
and Nicolas Vukadinovic\IEEEauthorrefmark{6}}
\IEEEauthorblockA{\IEEEauthorrefmark{1}LabSTICC, CNRS, Universit\'e de Bretagne Occidentale, 29238 Brest, France}
\IEEEauthorblockA{\IEEEauthorrefmark{2}SPEC, CEA, CNRS, Universit\'e Paris-Saclay, 91191 Gif-sur-Yvette, France}
\IEEEauthorblockA{\IEEEauthorrefmark{3}Plateforme technologique RMN-RPE, Universit\'e de Bretagne Occidentale, 29238 Brest, France}
\IEEEauthorblockA{\IEEEauthorrefmark{4}Univ. Grenoble Alpes, CEA, CNRS, Grenoble INP, INAC-Spintec, 38000 Grenoble, France}
\IEEEauthorblockA{\IEEEauthorrefmark{5}Institute of Physics, Kazan Federal University, Kazan 420008, Russian Federation}
\IEEEauthorblockA{\IEEEauthorrefmark{6}Dassault Aviation, 92552 Saint-Cloud, France}}

\maketitle

\begin{abstract}

  Liquid phase epitaxy of an 18~nm thick Yttrium Iron garnet (YIG)
  film is achieved. Its magnetic properties are investigated in the
  100 -- 400~K temperature range, as well as the influence of a 3~nm
  thick Pt overlayer on them. The saturation magnetization and the
  magnetocrystalline cubic anisotropy of the bare YIG film behave
  similarly to bulk YIG. A damping parameter of only a few $10^{-4}$
  is measured, together with a low inhomogeneous contribution to the
  ferromagnetic resonance linewidth. The magnetic relaxation increases
  upon decreasing temperature, which can be partly ascribed to
  impurity relaxation mechanisms. While it does not change its cubic
  anisotropy, the Pt capping strongly affects the uniaxial
  perpendicular anisotropy of the YIG film, in particular at low
  temperatures. The interfacial coupling in the YIG/Pt heterostructure
  is also revealed by an increase of the linewidth, which
  substantially grows by lowering the temperature.

\end{abstract}

\section{Introduction}

Yttrium Iron garnet (Y$_3$Fe$_5$O$_{12}$, YIG) is the marvel material
for ferromagnetic resonance (FMR) \cite{cherepanov93}, with the lowest
known Gilbert damping parameter, $\alpha = 3 \cdot 10^{-5}$ for bulk
\cite{spencer59,sparks64}. Since the 1970s, liquid phase epitaxy (LPE)
has been the reference method to grow micrometer-thick YIG films with
bulk-like dynamical properties \cite{henry73,wei84}, and numerous
microwave devices based on the propagation of spin-waves in such films
have been developed \cite{castera84}. In the past years, thin films of
YIG have become highly desirable in the context of magnonics
\cite{kruglyak10a,serga10} and its coupling to spintronics
\cite{chumak15} for three main reasons. Firstly, in magnonics, one
wishes magnetic films with low damping to ensure large propagation
length and with thickness limited to a few tens of nanometers so that
fundamental spin-wave modes do not interact with higher order
thickness modes, two conditions met in nanometer-thick YIG films
\cite{yu14}. Secondly, coupling YIG magnonics to spintronics was made
possible by the discovery that pure spin currents can be transmitted
at the interface between YIG (a magnetic insulator) and an adjacent
normal metal \cite{kajiwara10,cornelissen15}. In these hybrid
bilayers, ultrathin YIG layers are required to enhance the interfacial
effects. For instance, the damping of nanometer-thick YIG can be
controlled by the spin-orbit torque produced by an electrical current
flowing in an adjacent Pt layer \cite{hamadeh14b}, leading to the
generation of coherent spin-waves above a threshold instability
\cite{collet16}. Thirdly, due to the high resilience of YIG, only
nanometer-thick films can be patterned through standard
nanofabrication techniques, which is an important asset to design
magnonic crystals \cite{krawczyk14} and engineer the spin-wave
spectrum of individual YIG nanostructures \cite{hahn14}.

Due to this renew of interest for YIG thin films, lots of efforts have
been put in the last few years to produce ultrathin films with high
epitaxial and dynamical qualities. Pulsed laser deposition (PLD) is a
technique of choice to deposit ultrathin YIG layers ablated from
stoichiometric targets on Gd$_3$Ga$_5$O$_{12}$ (GGG) substrates
\cite{sun12,kelly13,onbasli14,chang14,lutsev16,hauser16}. Magnetic
properties close to bulk ones \cite{onbasli14} and Gilbert damping
parameters below $10^{-4}$ have been reported for films thinner than
50~nm \cite{hauser16}, even though such low intrinsic damping often
comes at the detriment of the full linewidth due to inhomogeneous
broadening \cite{chang14,lutsev16}. Another method to grow epitaxial
nanometer-thick YIG films is off-axis sputtering \cite{wang13}, which
can be used to control the strain-induced anisotropy on
lattice-mismatched substrates \cite{wang14c}. Finally, even though LPE
is not well suited to grow sub-micron thick films, it has been
successfully used to produce YIG films with thicknesses in the 100 --
200~nm range and damping parameters approaching $10^{-4}$
\cite{hahn13,dubs17}.

Here, we demonstrate that the growth of a YIG film as thin as 18~nm
and of high quality can be achieved by LPE. We present a detailed
investigation of its magnetic properties, including relaxation, as a
function of temperature, as well as the effect of a 3~nm thick Pt
overlayer on them.

\section{Sample preparation and preliminary characterizations at room
  temperature}

\begin{figure}
\centering
\includegraphics[width=7.4cm]{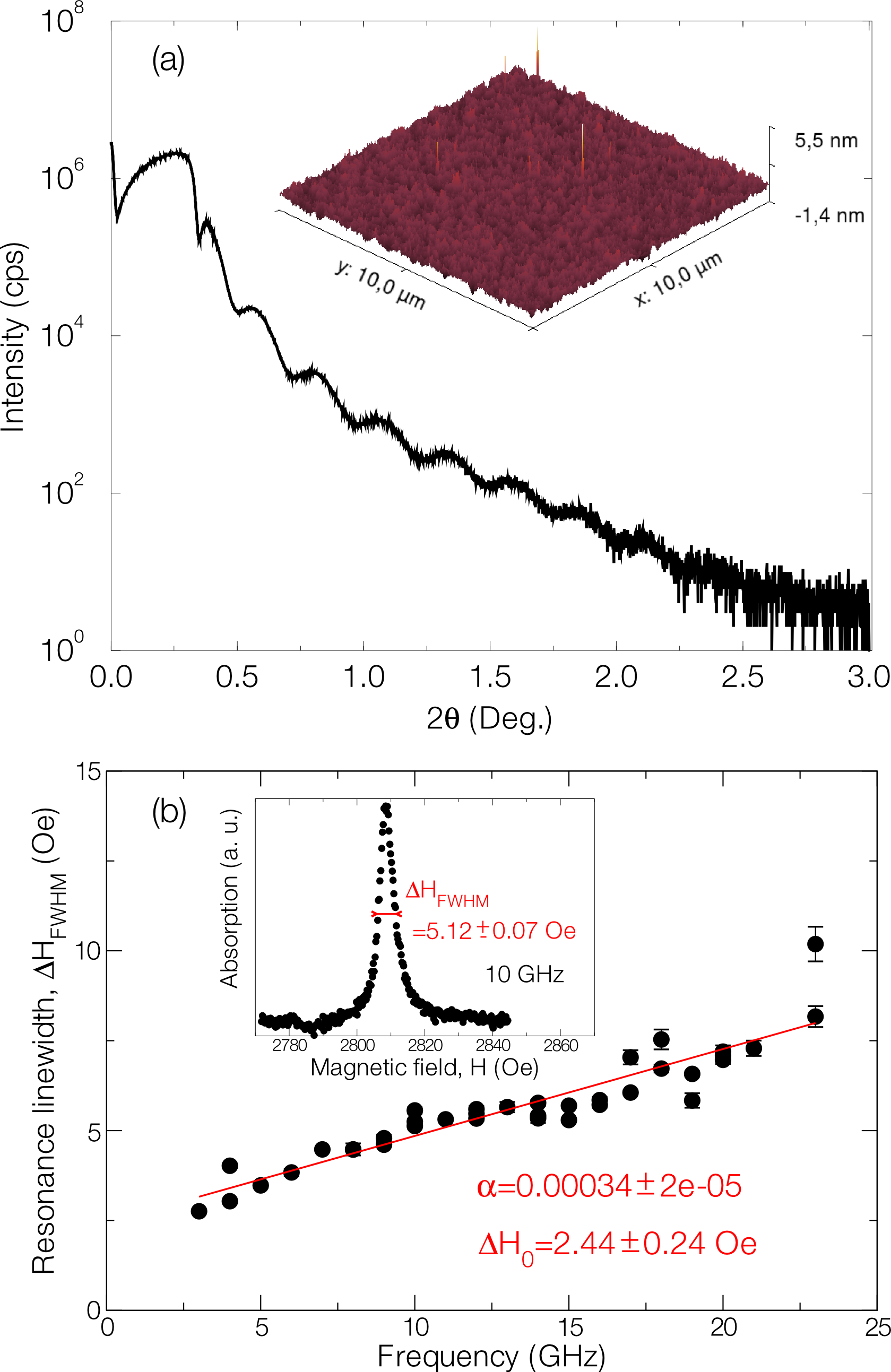}
\caption{(a) X-ray reflectometry of the 18~nm thick LPE grown YIG
  film. The inset shows the AFM surface topography. (b) Resonance
  linewidth \textit{vs.} frequency of the bare YIG layer measured by
  broadband FMR at room temperature. The inset displays the FMR line
  at 10~GHz.}
\label{Fig1}
\end{figure}

The ultrathin YIG film under consideration was grown by LPE from PbO
and B$_{2}$O$_{3}$ flux on a 1 inch (111)-oriented GGG
substrate. High-purity oxides (6N) were used. The key parameter is the
growth rate that depends on the growth temperature. To get an
ultrathin YIG film, a very low growth rate is required which means
that the growth temperature has to be close to the saturation
temperature (small supercooling). For the selected sample, the growth
rate was around 1~nm/s. Another important parameter is the speed of
the substrate rotation during the growth process that must be
controlled. The quality of the utrathin YIG film was checked by atomic
force microscopy (AFM) measurements from which a surface roughness of
about 0.25~nm was determined (see inset in Fig.1(a)). The film
thickness of 18~nm was confirmed by X-ray reflectivity measurements
(Fig.1(a)). From the experimental in-plane magnetization curves
recorded along the $[11{\overline{0}}]$ and $[11{\overline{2}}]$ axes
(not shown here), several conclusions can be drawn: (i) the saturation
magnetization $4{\pi}M_{S}$ at room temperature is equal to 1700~G,
close to the bulk value, (ii) the magnetization curves are isotropic
in the film plane, and (iii) the film has a very weak coercive field
($H_{c}{\simeq}0.3$~Oe).

Broadband FMR on a millimeter-size slab of the grown film was used to
investigate its magnetic relaxation at room temperature. The full
width at half maximum (FWHM) of the resonance line measured in
in-plane magnetized configuration as a function of the excitation
frequency is shown in Fig.1(b). The linear dependence of the linewidth
on frequency allows to fit a Gilbert damping parameter $\alpha = (3.4
\pm 0.2) \cdot 10^{-4}$ and an inhomogeneous contribution $\Delta H_0
= 2.44 \pm 0.24$~Oe. These values, which are close to those reported
on PLD grown YIG films of similar thickness \cite{kelly13}, highlight
the good dynamical quality of our film.

\section{Temperature dependence of magnetic properties for the bare
  YIG film}

\begin{figure}
\centering
\includegraphics[width=7.4cm]{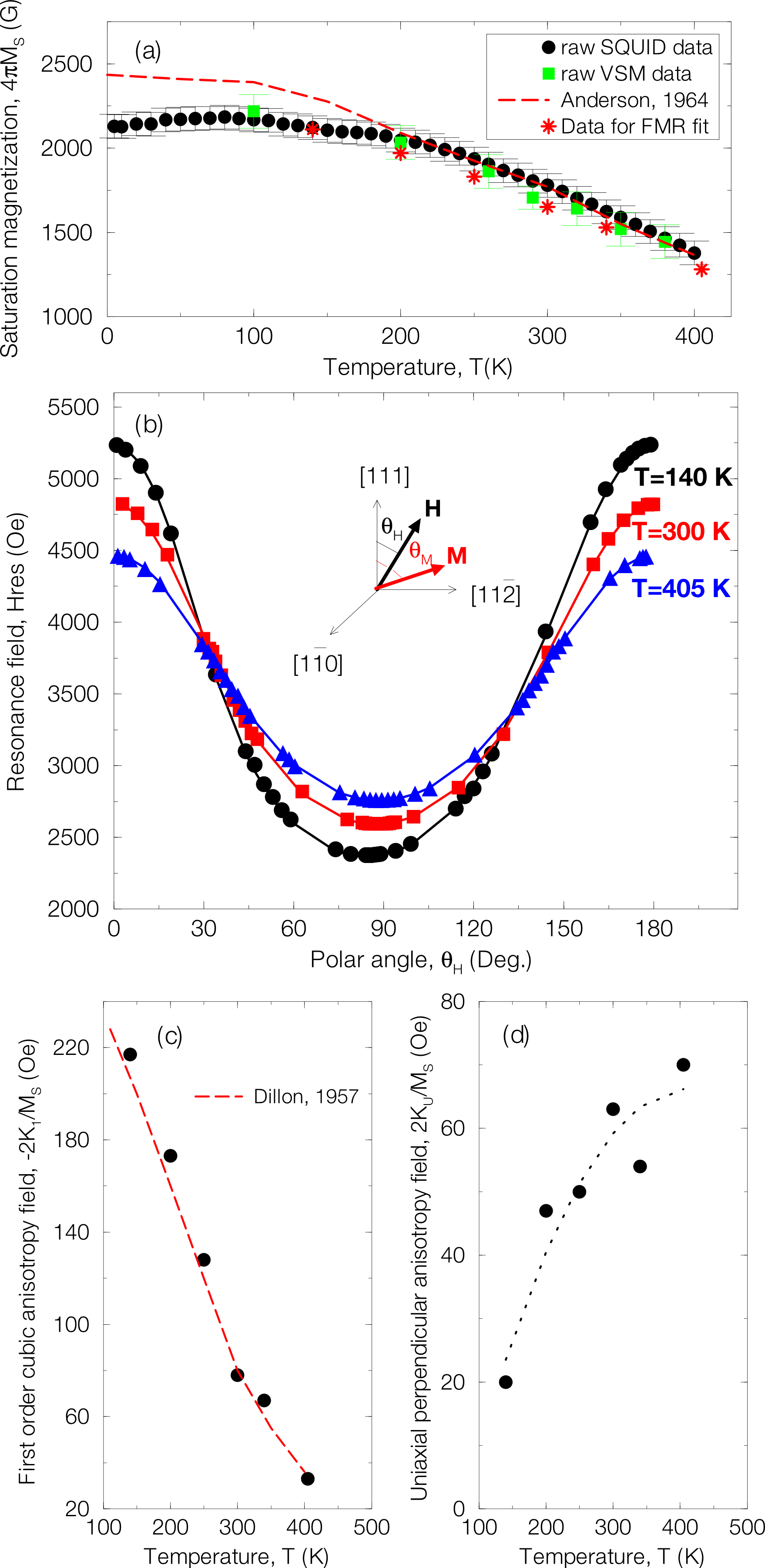}
\caption{(a) Dependence of the magnetization of the 18~nm thick YIG
  film on temperature. (b) Resonance field \textit{vs.} polar angle
  measured at X-band at three different temperatures. (c) Cubic and
  (d) uniaxial perpendicular anisotropies extracted as a function of
  temperature. The red dashed lines in (a) and (c) are the dependences
  for bulk YIG from literature, the dotted line in (d) is a guide to
  the eye.}
\label{Fig2}
\end{figure}

The temperature dependence of the saturation magnetization for the
bare YIG film measured by means of a superconducting quantum
interference device (SQUID) magnetometer and a vibrating sample
magnetometer (VSM) is displayed in Fig.2(a). The $4{\pi}M_{S}(T)$
profiles are consistent between the two methods and are in very good
agreement with the one reported for a bulk YIG sample
\cite{anderson64} for $T \ge 200$~K. For lower temperatures, the
saturation magnetization for the bulk sample exceeds the one found for
our ultrathin YIG film.

Next, the anisotropy constants were extracted from FMR measurements
performed in a X-band cavity (magnetic field swept at fixed frequency
$f=9.3$ GHz). We note that the slab used in this experiment has a
slightly larger inhomogeneous contribution to the linewidth than the
one measured by broadband FMR. Using the Makino's procedure
\cite{makino81}, the first-order cubic anisotropy constant $K_{1}$,
the first-order uniaxial perpendicular anisotropy constant $K_U$ and
the gyromagnetic ratio ${\gamma}$ were determined as a function of
temperature by exploiting the polar angle variation of the polarizing
magnetic field, the $4{\pi}M_{S}(T)$ profile being known for that very
same slab of film (red stars in Fig.2(a)). Examples of such variations
are reported in Fig.2(b) for three temperatures: $T=140$~K, 300~K, and
405~K, where ${\theta}_{H}$ is the polar angle of the polarizing
magnetic field. ${\theta}_{H}=0^\circ$ corresponds to the film normal
($[111]$ crystallographic axis) and ${\theta}_{H}=90^\circ$ is the
in-plane $[11{\overline{2}}]$ axis. Two remarks can be made. First,
the amplitude of the angular variation increases for decreasing
temperature. Second, this variation becomes more and more asymmetrical
with respect to the film plane as the temperature is lowered. This
last point means that the absolute value of $K_{1}$ is enhanced for
decreasing temperature. The experimental temperature dependence of
$K_{1}$ is displayed in Fig.2(c). The increase of the absolute value
of $K_{1}$ for decreasing $T$ satisfactorily matches with the
variation found for a bulk YIG sample \cite{dillon57}. On the other
hand, the temperature dependence of $K_U$ is reported in Fig.2(d). It
appears that the sign of $K_U$ is positive (easy axis along the film
normal) and $K_U$ increases with $T$.

The origin of $K_U$ in micrometer-thick LPE grown YIG films has been
discussed in detail in the eighties. Two main contributions were
identified, namely, the growth and the stress anisotropies. For pure
YIG films, the growth anisotropy is mainly induced by the Pb
impurities from solvent. It has been established that it raises with
the Pb content \cite{desvignes87}. In addition, the Pb content
increases with the supercooling \cite{hansen83}. In our case, the
growth of the ultrathin YIG film with a small supercooling prevents a
large contribution of the growth anisotropy. On the other hand, the
uniaxial stress anisotropy arises from the lattice parameter mismatch
between the GGG substrate ($a_{s}=12.383$~\AA) and the YIG film
($a_{f}=12.376 $~\AA). Introducing
${\Delta}a^{\perp}=(a_{s}-a_{f}^{\perp})/a_{s}$ where $a_{f}^{\perp}$
is the lattice parameter of the YIG film in the growth direction, the
uniaxial stress anisotropy constant $K_{U}^{s}$ is expressed by:
$K_{U}^{s}=(-3/2){\lambda}_{111}{\sigma}$ where
${\sigma}=E/(1-{\nu}){\Delta}a^{\perp}$, where $E$ is the Young's
modulus and ${\nu}$ the Poisson's ratio. For our film grown under
tension, ${\sigma}$ is positive, ${\lambda}_{111}$ is negative and a
positive value of $K_{U}^{s}$ is expected in agreement with the
experimental data. An estimate of $K_{U}^{s}$ based on the values of
$E$, ${\nu}$ and ${\lambda}_{111}$ for a bulk YIG sample leads to
about 40\% of the $K_{U}$ value deduced from FMR measurements. The
remaining difference can be ascribed to a surface induced contribution
to the uniaxial perpendicular anisotropy, which is also present in
ultrathin films. In fact, we have observed at room temperature that
$K_U$ varies with the YIG thickness, and becomes negative (easy plane)
above about 50~nm.

Moreover, a value of the gyromagnetic ratio nearly independent of
temperature was extracted from the angular fit of the resonance field,
${\vert}\gamma{\vert}=1.7685{\pm}0.002$~s$^{-1}$Oe$^{-1}$. This value
is consistent with the one deduced from the broadband FMR
measurements,
${\vert}\gamma{\vert}=1.771{\pm}0.0007$~s$^{-1}$Oe$^{-1}$.

\begin{figure}
\centering
\includegraphics[width=7.4cm]{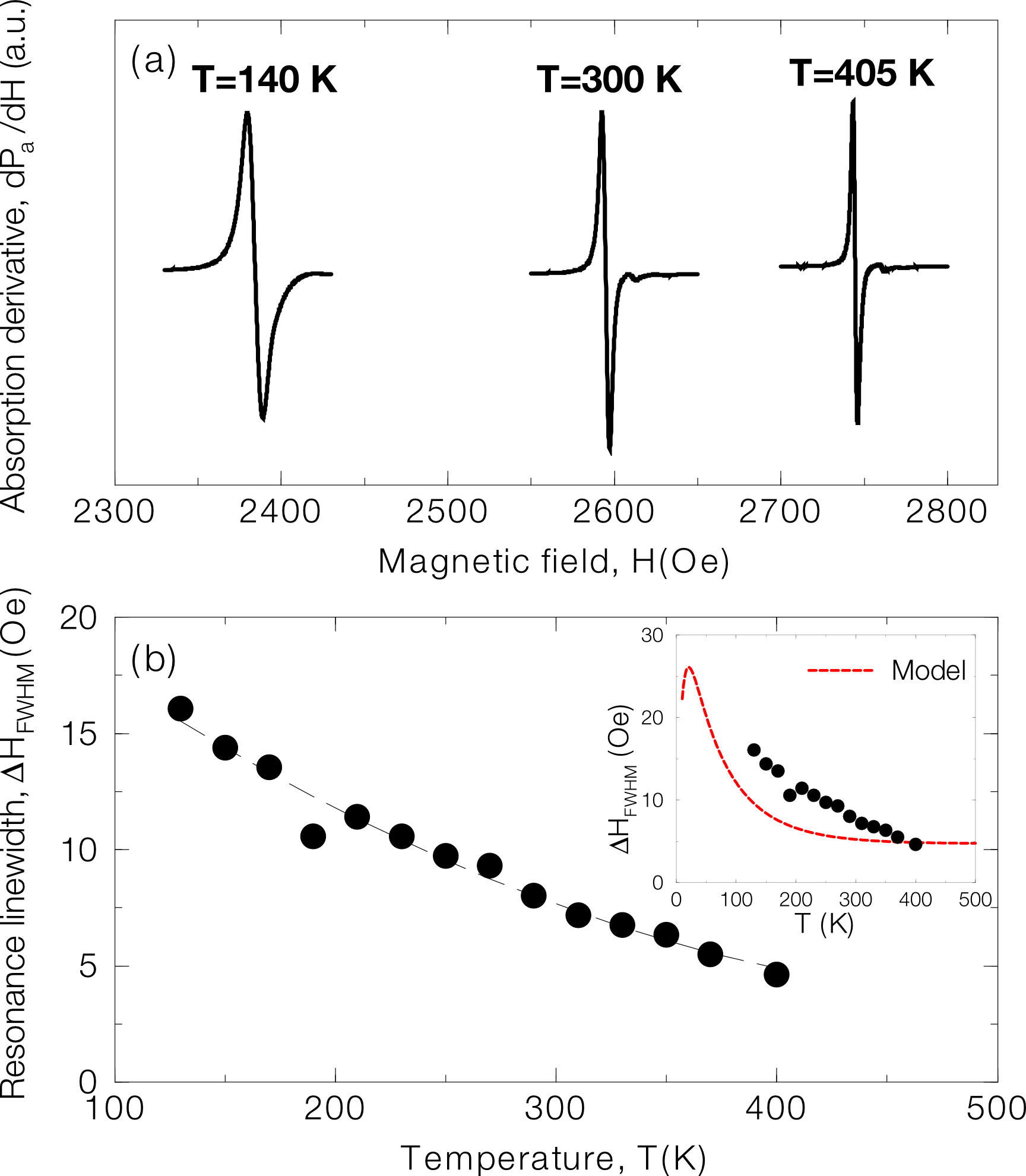}
\caption{(a) FMR lines measured at X-band on the bare YIG layer for
  three different temperatures. (b) FMR linewidth \textit{vs.}
  temperature. The inset shows the comparison of the data to the
  slowly relaxing impurity model (Eq.1).}
\label{Fig3}
\end{figure}

Another interesting feature is the temperature dependence of the
magnetic relaxation. Fig.3(a) shows the absorption spectra (derivative
form) versus magnetic field recorded at 9.3~GHz in the parallel
configuration for three temperatures. These spectra reveal that the
FMR linewidth increases for decreasing temperature. The value of
${\Delta}H_\text{FWHM}$ at 140~K is nearly four times greater than the
one at 405~K. Such a behavior has been recently reported on ultrathin
YIG films grown either by a spin coating method \cite{shigematsu16} or
by off-axis sputtering \cite{jermain17}, and on YIG spheres
\cite{maier-flaig17}. The broader temperature range probed in these
studies allows to evidence a peak-like maximum in the temperature
dependence of the FMR linewidth. The position of this peak is
frequency dependent and is located around $T=25$~K at X-band. Based on
these observations, the origin of the resonant temperature dependent
linewidth was ascribed to the slowly relaxing impurity mechanism
extensively investigated on bulk YIG samples in the sixties
\cite{spencer59,sparks64}. The existence of rare earth or Fe$^{2+}$
impurities induced during the growth process was put forward. In both
cases, the contribution of the slowly relaxing impurity mechanism to
the linewidth can be described by the following expression
\cite{jermain17}:
\begin{equation} \label{eq1}
\displaystyle
{\Delta}H_{SR}=A(T)\frac{{\omega}{\tau}}{1+({\omega}{\tau})^{2}} \, ,
\end{equation}
where $A(T)$ is a frequency-independent prefactor and ${\tau}$ a
temperature-dependent time constant. The total FMR linewidth
${\Delta}H_\text{FWHM}={\Delta}H_{0}+\frac{4{\pi}{\alpha}f}{{\vert}{\gamma}{\vert}}+{\Delta}H_{SR}$
is displayed in the inset in Fig.3(b) (red dashed curve). It can be
seen that the experimental temperature dependence of the FMR linewidth
cannot be fully reproduced by the model, where only ${\Delta}H_{SR}$
is a function of temperature. Other relaxation mechanisms seem to
contribute to the FMR linewidth. As highlighted in Ref.\cite{dubs17},
the existence of a transition layer between LPE grown YIG films and
the GGG substrate, whose thickness is estimated to be around 5~nm
(nearly a third of our film thickness), could produce additional
relaxation channels. For our ultrathin film, a sizeable surface
induced contribution to the linewidth can also be expected.

\section{Temperature dependence of magnetic properties for the YIG/Pt
  heterostructure}

\begin{figure}
\centering
\includegraphics[width=7.4cm]{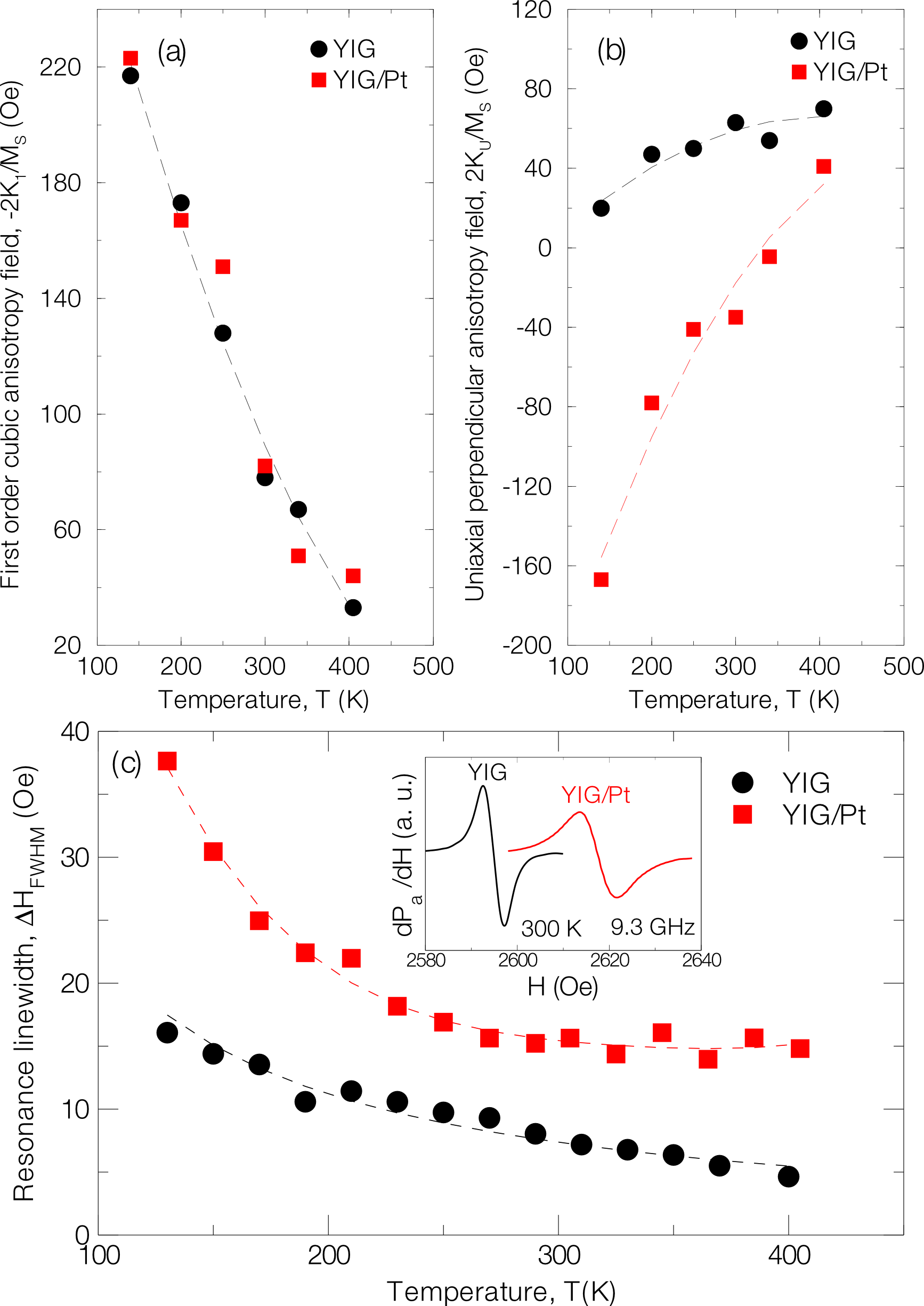}
\caption{Dependences on temperature of (a) cubic anisotropy, (b)
  uniaxial perpendicular anisotropy and (c) resonance linewidth for
  both the bare YIG layer (black dots) and the YIG layer capped with 3
  nm Pt (red squares). The inset shows the FMR lines measured at 300~K
  on both samples. All dashed lines are guides to the eye.}
\label{Fig4}
\end{figure}

Next, we are interested in the influence of a thin overlayer of Pt on
the magnetic properties of our ultrathin YIG film. Using electron beam
evaporation, a 3~nm thick Pt layer was evaporated on top of the YIG
film after a very soft \textit{in situ} dry etching of its
surface. Similarly to the bare YIG film, the obtained YIG/Pt
heterostructure was then studied at X-band as a function of
temperature. The main results of this characterization are compared to
those obtained on the bare YIG film in Fig.4. In order to extract the
anisotropy constants of the YIG/Pt bilayer, we used the same $4 \pi
M_S(T)$ profile as in Fig.2(a), since it is known that no magnetic
moment is induced in Pt by proximity effects at ferrites/Pt interfaces
\cite{collet17}. As can be seen in Fig.4(a), the Pt overlayer does not
change the cubic anisotropy of the bare YIG film, which is expected
from its bulk magnetocrystalline origin. In contrast, it does strongly
affect the uniaxial perpendicular anisotropy, which becomes negative
(easy plane) at low temperature, as shown in Fig.4(b). It means that
an interfacial mechanism \cite{tang18} of spin-orbit origin changes
the surface anisotropy component of $K_U$ discussed above for the bare
YIG film. The gyromagnetic ratio remains unaffected by the Pt
overlayer within our determination uncertainty.

As expected, the interfacial coupling between YIG and Pt also leads to
the increase of the FMR linewidth \cite{hurdequint07,heinrich11},
which is obvious from the comparison presented in Fig.4(c). At 300~K,
the measured linewidth is nearly doubled by the deposition of Pt on
top of YIG. To prove its interfacial origin, we have checked that this
enhancement is inversely proportional on the YIG thickness by varying
the latter in the 15 -- 200~nm range (not shown here). It is
interesting to note that the increase of the linewidth due to the Pt
overlayer is not constant as a function of temperature. The increment
in the damping depends both on the strength of the interfacial
coupling (whose microscopic origin is rarely discussed, most often, it
is parametrized by the so-called spin mixing conductance of the hybrid
interface) and on the spin transport parameters in the Pt layer (the
thickness of the latter being here comparable to the spin diffusion
length in Pt at room temperature \cite{castel12a}). These two
contributions can be temperature dependent, leading to a non trivial
behavior of the increment in the damping versus temperature. As a
matter of fact, the latter seems to be minimum around 250~K, and its
rapid increase below this temperature should be confirmed by
measurements at even lower temperature \cite{shigematsu16}.

\section{Conclusion}

In sum, this study demonstrates that the LPE growth method can be used
to produce YIG films with thickness below 20~nm and good structural
parameters. Surface induced contributions to the anisotropy and
relaxation are present in these ultrathin films. The effect of a Pt
overlayer proves that the interfacial coupling with this heavy metal
is efficient. In fact, similar batches of ultrathin LPE grown YIG
films have already been used to study the transport of angular
momentum in non-local YIG/Pt devices using spin-to-charge
interconversion \cite{thiery18,thiery18a}.

\section*{Acknowledgment}
We would like to thank Philippe Eli\`es and Fran\c cois Michaux from
the PIMM-DRX platform at UBO for their help with the AFM
characterization and the X-ray reflectometry.

\newpage


\end{document}